\documentclass[final,sort&compress,5p,twocolumn]{elsarticle}
\usepackage{epsfig}

\usepackage{amssymb}

\journal{NIM A}

\begin{document}

\begin{frontmatter}


\title{Physics with the ALICE Transition Radiation Detector}
\author{\mbox{Yvonne Pachmayer}\fnref{label2}}
\ead{pachmay@physi.uni-heidelberg.de}


\address[label2]{Physikalisches Institut, University of Heidelberg (Germany)}
\author{for the ALICE Collaboration}


\begin{abstract}

\noindent
The ALICE Transition Radiation Detector (TRD) significantly enlarges the scope of physics observables studied in ALICE, because it allows due to its electron identification capability to measure open heavy-flavour production and quarkonium states, which are essential probes to characterize the Quark-Gluon-Plasma created in nucleus-nucleus collisions at LHC. In addition the TRD enables to enhance rare probes due to its trigger contributions. \newline
\noindent We report on the first results of the electron identification capability of the ALICE Transition Radiation Detector (TRD) in pp collisions at $\sqrt{s}$ = 7 TeV using a one-dimensional likelihood method on integrated charge measured in each TRD chamber. The analysis of heavy flavour production in pp collisions at $\sqrt{s}$ = 7 TeV with this particle identification method, which extends the $p_{\rm t}$ range of the existing measurement from $p_{\rm t}$ = 4 GeV/$c$ to 10 GeV/$c$ and reduces the systematic uncertainty due to particle identification, is presented. The performance of the application of the TRD electron identification in the context of $\rm J/\psi$ measurements in Pb-Pb collisions is also shown.

\end{abstract}


\end{frontmatter}


\section{Introduction}\label{Section_Intro}

\noindent A Large Ion Collider Experiment (ALICE) \cite{ALICEexperiment} is the dedi\-cated detector setup to study all aspects of heavy ion
collisions at the LHC at CERN. It is assumed that in nucleus-nucleus collisions at high energies a 
high density de-confined state of strongly interacting matter, known as Quark-Gluon-Plasma (QGP), is created. \mbox{ALICE} is designed to measure a large set of observables in order to study the properties of this QGP. The Transition Radiation Detector (TRD) \cite{ALICETRD} provides electron identification in the ALICE central barrel ($|\eta|$ $\rm < 0.9$) at momenta \mbox{$p \rm >$ 1 GeV/$c$}, where pions cannot be rejected anymore sufficiently via energy loss measurements in the Time Projection Chamber (TPC). In addition the TRD can be used to trigger on identified particles with high transverse momenta, thus providing enriched samples of single electrons or electron-positron pairs, and to select jets. Therefore the TRD significantly enlarges the scope of physics observables.\newline
\noindent The following probes are regarded as essential probes of the conditions of the Quark-Gluon-Plasma and can be accessed with the help of the TRD:

\begin{figure*}[htb]
      \hspace{1.0cm}
     \begin{minipage}{0.35\textwidth}
      \centering
       \includegraphics[width=1\textwidth]{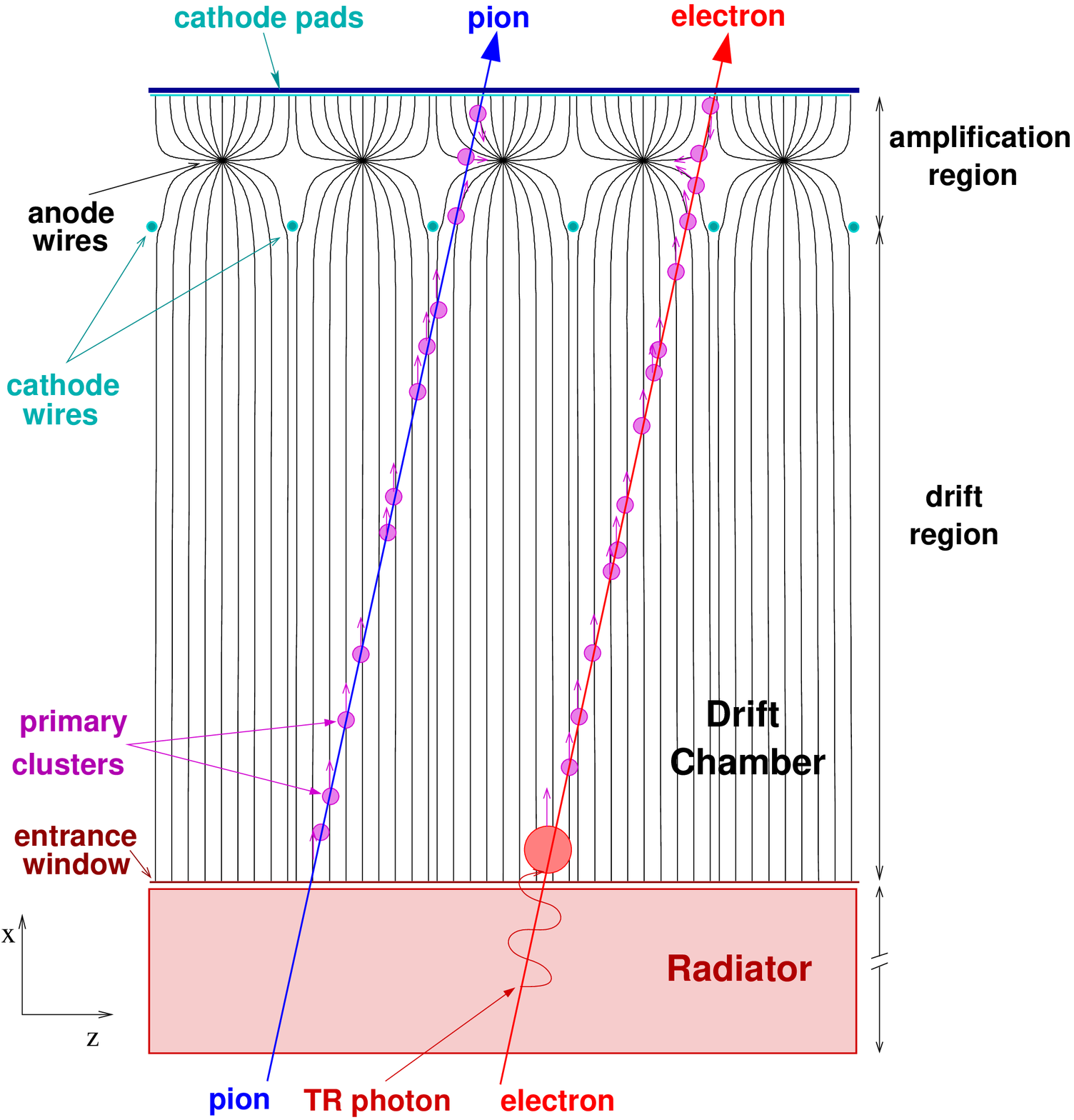}
       \caption{Schematic cross-section of a TRD chamber including radiator.}\label{workingprinciple}
     \end{minipage}
     \hspace{2.5cm}
     \begin{minipage}{0.35\textwidth}
      \centering
    \includegraphics[width=1\textwidth]{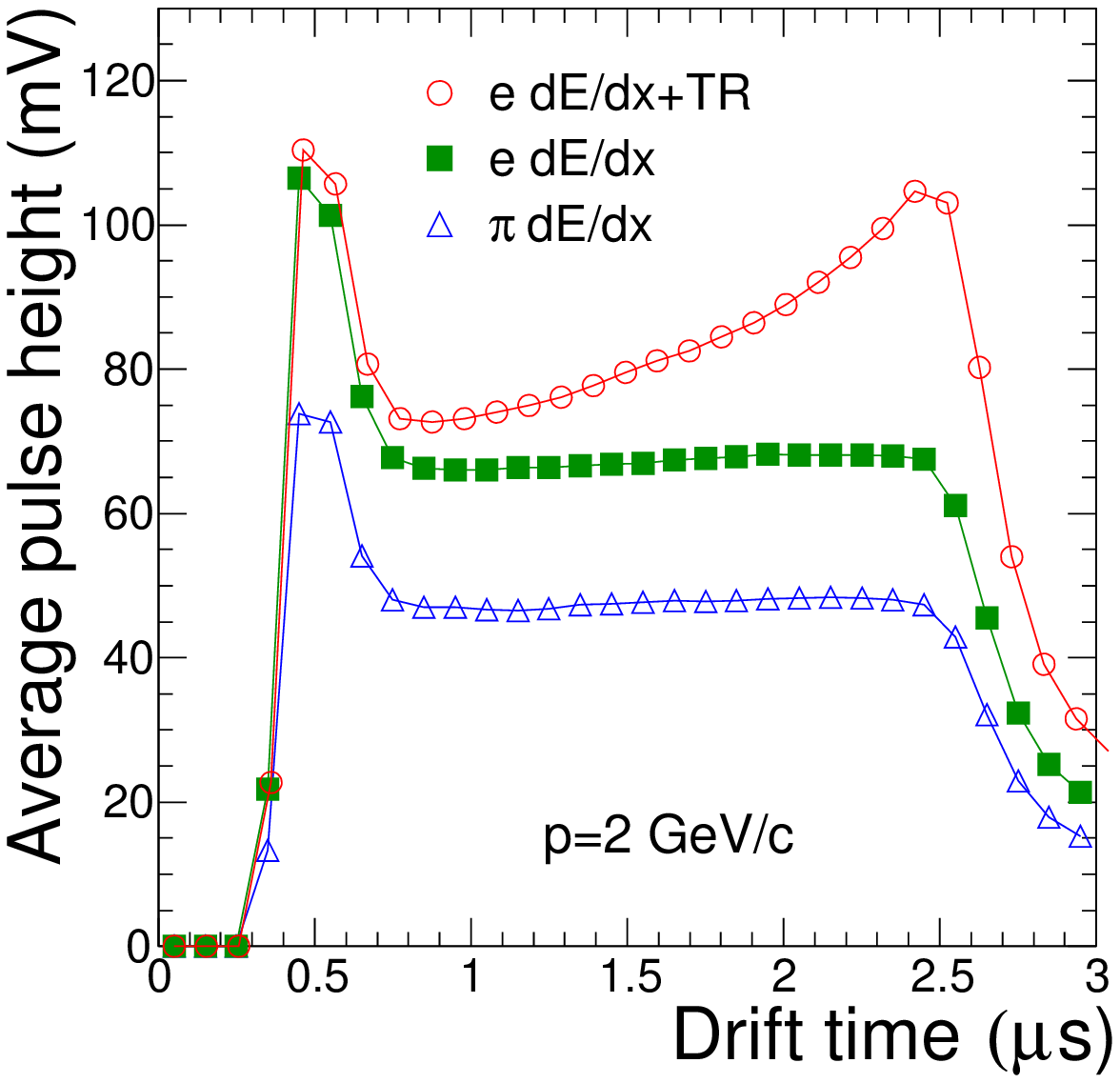} 
       \caption{Average pulse height as a function of drift time for pions and electrons with and without radiator (testbeam measurements) \cite{Anton}.}\label{PHfigure}
     \end{minipage}
   \end{figure*}

\begin{itemize}

\item Open heavy-flavour production \newline
Open charm and beauty can be detected with the \mbox{ALICE} central barrel in the semi-electronic decay channel: D, B $\rightarrow$ ${\rm e}^{\pm}$ + X .
\noindent Because of their high mass charm and beauty quarks are mainly produced on a very short time scale ($\tau$ $\approx$ 1/(2$m_{\rm q}$) $\lesssim$ 0.1 fm/$c$) in hard scattering processes during the initial stage of the collision. Thus heavy quarks witness the entire evolution of the dense matter and are 
sensitive to the properties (gluon densities, temperature, volume) of the de-confined medium through
in-medium energy loss. Open charm and beauty production cross sections are also needed as reference for quarkonia studies, because at LHC energies heavy quarks are mainly produced through gluon-gluon fusion processes $\rm (gg \rightarrow Q\overline{Q})$. In addition the B meson production measurement allows to estimate the contribution of secondary $\rm J/\psi$ $\rm (B \rightarrow J/\psi + X)$ to the total $\rm J/\psi$ yield.

\item Quarkonium states ($\rm c\overline{c}$ and $\rm b\overline{b}$) \newline
\noindent $\rm J/\psi, \psi'$ and $\rm \Upsilon$ can be measured with the ALICE central barrel in the di-electron channel. According to one prediction the $\rm c\overline{c}$ and $\rm b\overline{b}$ states should be sensitive to the temperature of the QGP, because of their dissociation due to colour-screening \cite{Satz}.
Charmonium regeneration models \cite{PBM, Thews} on the other hand predict that $\rm c\overline{c}$ bound states can be abundantly produced by recombination of $\rm c$ and $\rm \overline{c}$ quarks during hadronization. This recombination mechanism would lead to a quarkonium enhancement at high energies. 

\item Jets and $\rm \gamma$-jet production \newline
\noindent These probes allow to study parton energy loss and medium-modified fragmentation functions to unravel the detailed spatial and temporal structure of the QGP. The TRD provides the necessary trigger to enhance the statistics of these data samples.

\item Virtual photons $\rm \rightarrow e^{+}e^{-}$ and Drell-Yan production \newline
\noindent These processes are very challenging to detect. However they allow to investigate thermal radiation in the mass window between $\rm J/\psi$ and $\rm \Upsilon$. Thermal radiation not only serves as a proof of de-confinement, but also provides a direct measurement of the temperature of the QGP.
\end{itemize}

\noindent For reference the corresponding studies have to be performed in proton-proton collisions. In addition, e.g. the measurements of heavy-flavour production in pp collisions allow to test perturbative Quantum Chromodynamics in a new energy domain. In the case of hadroproduction of quarkonium states non-perturbative aspects are also involved. The ALICE physics program further covers \mbox{studies} of p-Pb collisions in order to disentangle initial and final state effects.

\section{The Transition Radiation Detector}\label{Section_TRD}
\noindent To achieve the physics goals listed above the ALICE TRD was designed to reject pions by a factor 100, to work in the high multiplicity environment of heavy ion collisions as well as to provide in the bending plane a space point resolution better than \mbox{400 $\rm \mu$m} and an angular resolution better \mbox{than 1$\rm ^{\circ}$.} \newline
\noindent The TRD consists of 522 chambers arranged in 6 layers surrounding the TPC at a radial distance r (2.9 $\leq$ r \mbox{$\leq$ 3.7 m)} from the beam axis, with a maximum length of 7 m along the beam axis (corresponding to a pseudo-rapidity coverage of $|\eta|$ $\rm < 0.9$). It has an active area of roughly \mbox{675 $\rm m^{2}$} and the total radiation length for the six layers is about \mbox{25\% of $\rm X_{0}$}.  \mbox{Figure \ref{workingprinciple}} shows the cross-section of a TRD chamber, which has an average size\footnote{The dimensions change with distance from the interaction point to achieve a constant granularity.} of \mbox{135 cm $\times$ 103 cm} and is about 12 cm thick, including radiator, electronics and cooling. The gas volume, filled with a $\rm Xe~CO_{2}$ mixture (85:15), is subdivided by a cathode wire grid into a 3 cm drift region and 0.7 cm amplification region equipped with anode wires. The signal induced on the segmented cathode plane is typically spread over 2-3 pads. Each pad has an average size\footnotemark[\value{footnote}] of 0.7 cm $\times$ 8.7 cm. The front-end-electronics is directly mounted on the back of the TRD chamber. The pad signals are read out and processed in a custom built charge sensitive preamplifier-shaper circuit and digitized by a 10 MHz ADC to record the time evolution of the signal. The drift time of about \mbox{2 $\mu$s} is sampled in 20 time bins. \newline
\noindent Figure \ref{PHfigure} shows the average pulse height as a function of drift time for pions and electrons with a momentum of \mbox{2 GeV/$c$} passing through the TRD chamber. The peak at early drift times is due to the charge deposit in the amplification region, where the charge drifts from both sides to the anode wire. The plateau at intermediate drift times originates from the drift region. Comparing pions with electrons one sees that the latter deposit on average more charge, because of the higher specific energy loss at this momentum. \newline \noindent A 4.8 cm thick radiator, a sandwich of polypropylene fibers and Rohacell foam, which provides many boundaries between media with different dielectric constants, is placed in front of each drift region. \newline
\noindent Transition Radiation (TR) is emitted if a relativistic particle ($\gamma \rm \gtrsim$ 1000) traverses the radiator \cite{XLu}. At LHC energies TR is only generated by electrons. The produced TR photons (1-30 keV) are absorbed in the drift volume preferentially close to the radiator by the high Z-gas mixture (Xe-based), which corresponds to long drift times to the anode wire. This contribution is visible in Fig. \ref{PHfigure} as the characteristic peak at large drift times and enables an even better separation of electrons and pions.

\section{Electron Identification}\label{Section_PID}

\begin{figure}[htb]
   \centering
\includegraphics[width=.5\textwidth]{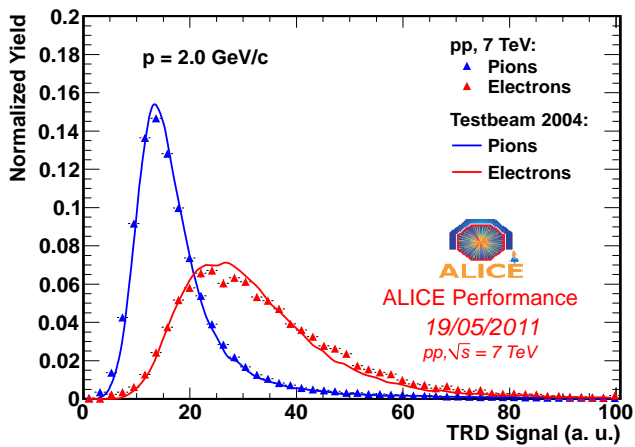} 
  \caption[]{Charge deposit per tracklet for electrons and pions in pp collisions compared with testbeam measurements. Topological cuts were applied in pp collisions to select electrons from gamma-conversions and pions from $\rm K^{0}_{s}$ decays.}
   \label{Figure_trackletcharge}
\end{figure}

\noindent The track-by-track electron identification is at the moment based on the simplest method of TRD particle identification - one-dimensional likelihood on the total integrated charge measured in a single TRD chamber (tracklet). \newline
\noindent Figure \ref{Figure_trackletcharge} shows the charge deposit per TRD chamber for pions and electrons at a momentum of 2 GeV/c from pp collisions at $\sqrt{s}$ = 7 TeV compared with the testbeam measurements taken in 2004 \cite{Anton}. The distributions from pp collisions are in good agreement with the testbeam measurements. The average charge deposit is larger for electrons than for pions due to the higher specific energy loss and Transition Radiation. 

\begin{figure}[h]
   \centering
\includegraphics[width=.5\textwidth]{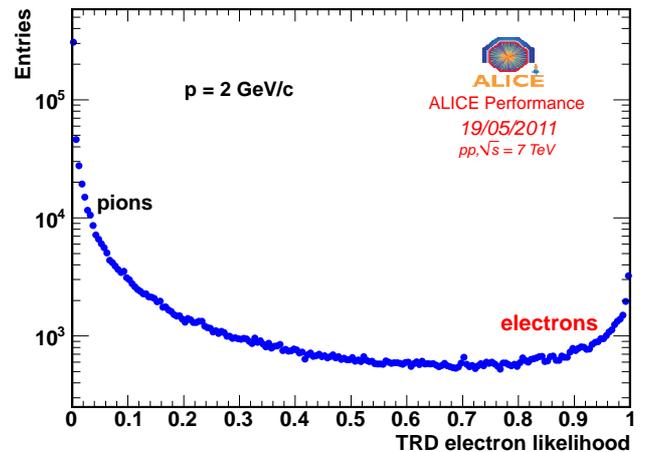} 
  \caption[]{Electron likelihood distribution after TOF particle identification for tracks with $p$ = 2 GeV/$c$ and reconstructed from at least 5 tracklets in pp collisions at $\sqrt{s}$ = 7 TeV. Since pions are much more abundant than electrons, the electron signal around 1 is on top of the tail of the pion distribution \cite{MFasel}.}
   \label{Figure_likeslice}
\end{figure}

\noindent During the testbeam time 2004 charge deposit distributions were measured for electrons and pions for the momenta 1, 1.5, 2, 3 to 10 GeV/c. These distributions are now used as reference (lookup table) for particle identification. Reference distributions for muons, kaons and protons were obtained via parameterizations from GEANT3 taking into account that because of their higher mass the deposited energy inside the detector is only due to ionization. The measured signal is then normalized to the testbeam references by comparing the charge deposit distribution for pions with $p$ = 1 GeV/c in testbeam with the one derived from pions originating from $\rm K^{0}_{s}$-decays (in pp collisions). The probability for a single tracklet (TRD chamber) of being an electron, respectively a pion, muon, kaon or proton is then determined via interpolation between adjacent momentum references. The probability for the full track, reconstructed from at least 4 TRD chambers (layers), is then calculated via Bayes' theorem. In case the Time-of-Flight system (TOF) is used beforehand for particle identification, the probabilites of being only an electron or pion are renormalized, because the TOF effectively rejects kaons and protons at low momenta. \newline
\noindent The likelihood distribution of all selected tracks after application of an electron hypothesis cut in TOF, i.e. the difference between expected and measured time of flight has to be less than 3$\sigma$, is shown in Fig. \ref{Figure_likeslice} for pp collisions at $\sqrt{s}$ = 7 TeV. To ensure that a defined constant electron efficiency is fulfilled for particle identification a momentum dependent cut on the likelihood is applied. This cut is tuned using electron candidates from conversion, which were selected using topological cuts. 
The threshold, which depends on the momentum as well as selected electron efficiency and number of tracklets, was then parameterized for usage in the analysis.

\begin{figure}[htb]
   \centering
\includegraphics[width=.5\textwidth]{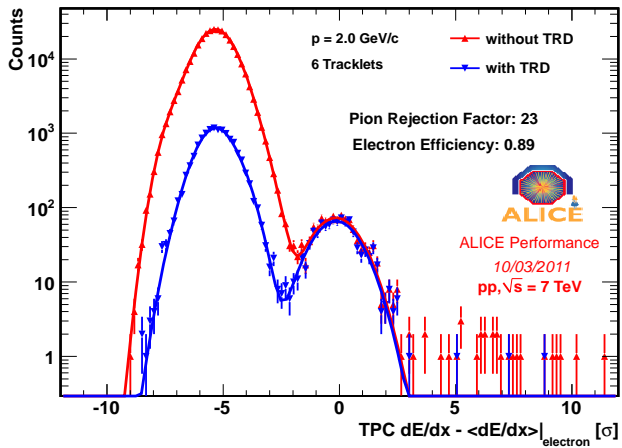} 
  \caption[]{TPC dE/dx signal (in units of standard deviations) relative to the electron Bethe-Bloch line for 2 GeV/$c$ tracks and 6 TRD layers, with and without TRD \cite{MFasel}.}
   \label{tpc_before_after_trd}
\end{figure}

\noindent Figure \ref{tpc_before_after_trd} visualizes the effect of the TRD with this simplest method of particle identification. Shown is the difference in units of standard deviations between the measured TPC energy loss of a given track and the calculated energy loss of an electron for tracks, reconstructed with 6 tracklets and momenta of 2 GeV/$c$, before and after applying the TRD electron identification. While 90\% of electrons are kept, pions are surpressed by a \mbox{factor 23}. 

\noindent Table \ref{tab1} summarizes the pion rejection factors for an electron efficiency of 80\% and 90\% for different numbers of layers. With decreasing electron efficiency and/or \mbox{increasing} number of tracklets the pion rejection factor increases as expected. \newline
\noindent Cross-checks with electrons from conversion show that the electron identification with this method is well under control. The electron likelihood is around 1 for this data sample and the selected electron efficiency is constant as a function of momentum. In addition, the tracking efficiency is in agreement with the one obtained from Monte Carlo (MC) simulations, which is important in view of correcting the experimental data. 

\noindent More powerful particle identification methods with the TRD, e.g. two-dimensional likelihood and neural networks, are under development. These methods make use of the uniqueness of the ALICE TRD, i.e. the recorded time evolution of the signal, where at large drift times the TR peak is measured. Studies show that for the two mentioned methods a pion rejection of better than 100 is expected for an electron efficiency of 90\% in the momentum range \mbox{1-2 GeV/$c$} (with 6 layers) \cite{Anton, RB, Wilk}.

\begin{table}[htb]
  \begin{center}
  \begin{tabular}{ c c c c}
  \hline
~Electron Efficiency~ & ~4 layers~  & ~5 layers~  & ~6 layers~  \\
  \hline
  90\%  & 9 & 15 & 23\\
  80\%  & 7 & 19 & 44\\
  \hline
  \end{tabular}
  \end{center}
   \caption[]{Pion rejection factor for an electron efficiency of 80\% and 90\% for different numbers of layers.}\label{tab1}
\end{table}

\section{Physics Results}\label{Section_Physics}

\subsection{Semi-leptonic Decays of Heavy-Flavour Hadrons}\label{Section_HFE}

\begin{figure}[htb]
   \centering
\includegraphics[width=.45\textwidth]{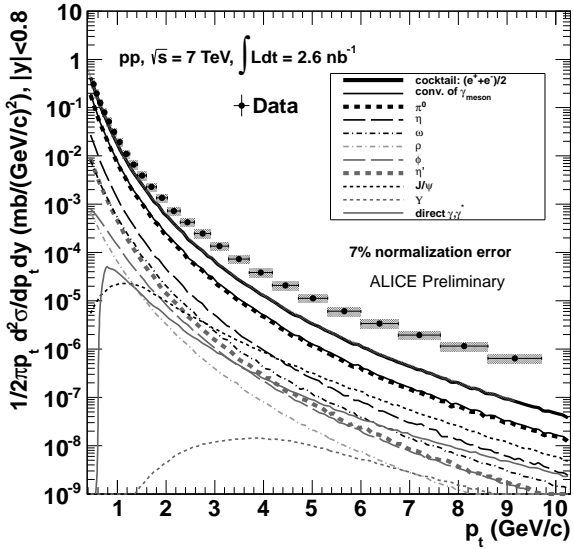} 
  \caption[]{Inclusive electron spectrum compared with the cocktail of electrons from background sources.}
   \label{Figure_Cocktail}
\end{figure}

\noindent The one-dimensional likelihood method to select electrons, described in Section \ref{Section_PID}, was applied for the first time in the semi-leptonic heavy-flavour analysis. \newline
\noindent The findings presented in this contribution are obtained from the 2010 pp run at \mbox{$\sqrt{s}$ = 7 TeV}. A total of about \mbox{$\rm 180 \cdot 10^{6}$ } minimum bias events ($\rm \cal{L}$ $_{\rm int} = 2.6~nb^{-1}$) were analyzed. \newline
\noindent In the pseudo-rapidity range $|\eta|$ $\rm < 0.9$ the ALICE detector system allows excellent track reconstruction and particle identification for electrons, especially because of the TRD.\newline
\noindent In the present analysis the Inner Tracking System (ITS) and the TPC were used for vertex and track reconstruction. After selecting good-quality tracks in the TPC, we required a hit in the innermost layer of the ITS \mbox{(r = 3.9 cm)} to reduce the background from photon conversions. The measured time of flight had to be consistent with the electron hypothesis within 3 $\sigma$ to reject kaons up to \mbox{$p$ = 1.5 GeV/$c$} and protons up to \mbox{$p$ = 3 GeV/$c$}. Next the one-dimensional TRD particle identification method was applied with an electron efficiency of 80\% for tracks reconstructed in at least 5 layers. Furthermore, a cut on the specific energy deposit d$E$/d$x$ in the TPC, expressed in number of sigmas from the electron line, was applied. This cut was set to 0 to 3 $\sigma_{\rm e}$. The remaining hadron contamination was estimated from fits of the TPC d$E$/d$x$ in momentum slices, and the yield was subtracted from the electron spectrum. Over the momentum range \mbox{0.5-10 GeV/$c$} the hadron contamination does not exceed \mbox{5\%}. \newline
\noindent The efficiency and acceptance corrections were derived from MC simulations using PYTHIA, combined with detector simulation and event reconstruction. Wherever possible the efficiency corrections were cross-checked with a data-driven method in which the signal from \mbox{$\rm \gamma\rightarrow$ e$^{+}$e$^{-}$} conversions in material was evaluated. The track reconstruction efficiency and the acceptance\footnote{At the time the TRD only covered 7/18 of the azimuthal angle.} of tracks reconstructed in at least 5 layers in the region \mbox{$|\eta|$ $\rm < 0.8$} reach about 20\% at $p_{\rm t} \sim$ 1.5 GeV/$c$. The resulting transverse momentum spectrum for inclusive electrons is shown in \mbox{Fig. \ref{Figure_Cocktail}}. \newline
\noindent To estimate the systematic uncertainty due to TRD track matching and particle identification the respective selection criteria were varied, i.e. the electron efficiency to 70\% for tracks with at least 4 layers and to 90\% for tracks with at least 6 layers (lower and upper maximum deviation). A momentum independent maximum uncertainty of 10\% was found. The total systematic uncertainty calculated as quadratic sum of all error sources (incl. ITS, TPC etc) within the analysis is 20\%. \newline
\noindent 

\noindent The inclusive electron yield consists primarily of three components: (i) electrons from heavy-flavour hadron decays, (ii) background electrons from Dalitz decays of light mesons ($\rm \pi^{0}$, $\rm \eta$, etc) and photon conversions as well as real and virtual direct photons and (iii) non-photonic background from di-electron decays of vector mesons and $\rm K_{\rm e3}$ decays. The $\rm 2^{nd}$ component is the largest background contribution. To extract the electrons from heavy-flavour hadron decays the background sources are subtracted with the so-called "cocktail method". Using a MC event generator of hadron decays a cocktail of electron spectra of background sources is calculated based on the cross section and the momentum distributions of the electron sources. \mbox{Figure \ref{Figure_Cocktail}} shows the inclusive electron spectrum compared with the corresponding cocktail of electrons from background sources. Electrons from $\rm \pi^{0}$-Dalitz decays \mbox{($\rm \pi^{0} \rightarrow$ $\rm \gamma$ e$^{\rm +}$e$^{\rm -}$)} and from photon conversions from the decay $\rm \pi^{0} \rightarrow \rm \gamma \rm \gamma$ in the detector material contribute dominantly to the background. Thus an excellent parametrization of the pion spectrum, both in absolute yield as well as in $p_{\rm t}$ shape, is mandatory. The pion input is based on $\rm \pi^{0}$ measurements with ALICE. The ratio of conversion to Dalitz decay is obtained from the known material budget. The heavier mesons are implemented via $m_{\rm T}$ scaling. The di-electron decays of   $\rm J/\psi$ and $\rm \Upsilon$ are included based on measurements with ALICE and CMS \cite{ALICEjpsi, CMSupsilon}. The QCD photons (direct $\rm \gamma, \rm \gamma^{*}$) are based on NLO calculations \cite{WVogelsang}. The systematic uncertainty of the cocktail is estimated to be 20\%. With increasing momentum the signal to background ratio increases. 

\begin{figure}[htb]
   \centering
\includegraphics[width=.5\textwidth]{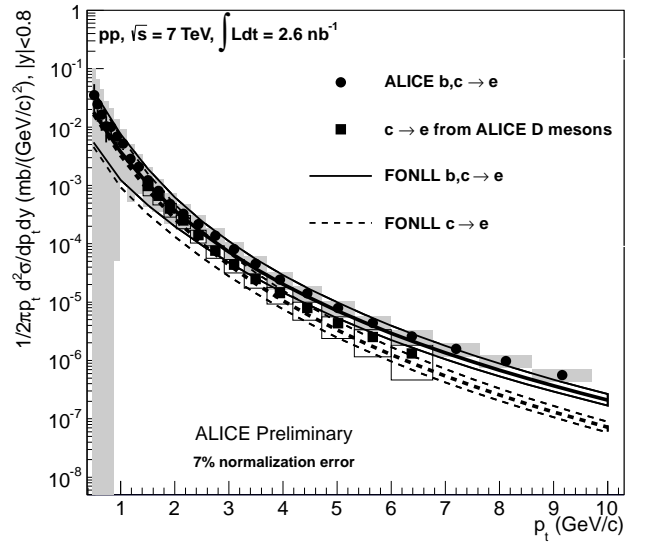} 
  \caption[]{Heavy-flavour decay electrons in $|\eta|$ $\rm < 0.8$ compared with FONLL calculations and electrons from charm decays derived by applying the PYTHIA decay kinematics to the D meson cross sections.}
   \label{Figure_HFEspectrum}
\end{figure}

\noindent The \mbox{$p_{\rm t}$ differential} production cross section of electrons from heavy-flavour hadron decays, shown in Fig. \ref{Figure_HFEspectrum}, is then extracted by subtracting the above described  cocktail of background electrons from the inclusive electron spectrum. The distribution is plotted in comparison with FONLL calculations \cite{FONLLcalc} and electrons from charm decays derived by applying the PYTHIA decay kinematics to the D meson cross sections \cite{ALICEdmeson}. At low $p_{\rm t}$, where the charm contribution dominates, both measurements are consistent. The FONLL calculations are in agreement with the data.  \newline
\noindent The advantage of the usage of the TRD in this analysis in comparison with an analysis only based on TOF and TPC results in a reduction of systematic uncertainty in particle identification
from 20\% to 10\% and an extended $p_{\rm t}$ range from 4 GeV/$c$ to 10 GeV/$c$. With the TRD, electrons can be further separated from pions at low momenta and above \mbox{$p$ = 4 GeV/$c$} electrons can only be clearly identified with the TRD. The present $p_{\rm t}$ reach is now only limited due to the currently analyzed statistics.

\subsection{$\rm J/\psi$ Meson}\label{Section_Quarkonia}
\begin{figure}[htb]
   \centering
\includegraphics[width=.42\textwidth]{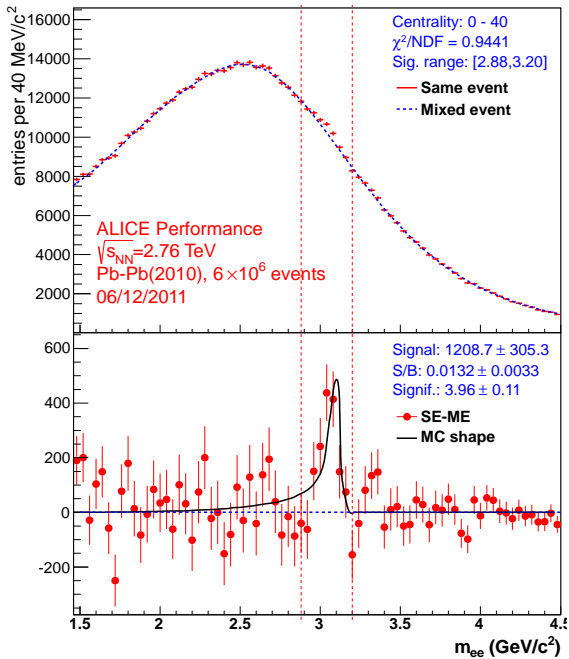} 
  \caption[]{$\rm e^+e^-$ invariant-mass distribution with only TPC particle identification for the centrality \mbox{0-40\%}.}
   \label{Figure_invmassnotrd}
\end{figure}

\begin{figure}[htb]
  \centering
\includegraphics[width=.42\textwidth]{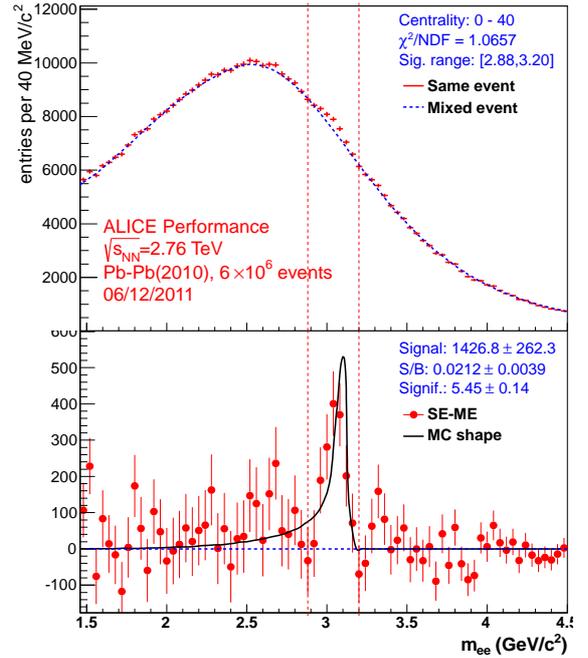} 
  \caption[]{$\rm e^+e^-$ invariant-mass distribution with TPC and TRD particle identification for the centrality \mbox{0-40\%}.}
   \label{Figure_invmasstrd}
\end{figure}

\noindent The performance of the $\rm J/\psi$ measurement with the TRD was studied in the first Pb-Pb run at \mbox{$\sqrt{s_{\rm NN}}=2.76$ TeV} taken in 2010. The results in this contribution show the present status of the ongoing analysis. \newline \noindent The Silicon Pixel Detector (SPD) at mid-rapidity and the forward VZERO scintillator counters provide a minimum bias trigger (corresponding to 97\% of the inelastic Pb-Pb cross section) and are also used to derive the centrality of the \mbox{collisions \cite{AToia}}. Events with a centrality of \mbox{0-40\%} were selected. The inclusive $\rm J/\psi$ measurement was based on the invariant mass of two identified electrons within the pseudo-rapidity range \mbox{$|\eta|$ $\rm < 0.9$}. The background was calculated with mixed events technique. A scaling to the same event distribution was performed for the invariant mass region \mbox{3.2 - 4.0 GeV/$c^{\rm2}$}. \newline
\noindent The invariant mass of electron pairs is shown in Fig. \ref{Figure_invmassnotrd}. The electrons were selected if their specific energy deposit d$E$/d$x$ in the TPC expressed in number of standard deviations from the electron expectation ranged within $\rm -2 < \sigma_{\rm e} < 3$. Furthermore $\rm \pm 3.5\sigma$ exclusion cuts for pions and protons were employed. After background subtraction one finds 1208 $\rm \pm$ 305
$\rm J/\psi$ candidates extracted by bin counting in the invariant mass region 2.88-3.20 GeV/$c^{2}$. 
The signal to background ratio is $\rm 0.0132 \pm 0.0033$ and the significance is $\rm 3.96 \pm 0.11$.\newline
\noindent Figure \ref{Figure_invmasstrd} shows the improvement when in addition the TRD electron identification was applied with an electron efficiency of 90\% for tracks reconstructed with at least 4 tracklets. Due to the strong discrimination power of the TRD the combinatorial background is reduced in comparison with the background in the analysis based only on the TPC. This becomes obvious from the larger signal to background ratio of 0.0212 $\rm \pm$ 0.0039 and the larger sig\-ni\-fic\-ance of \mbox{5.45 $\rm \pm$ 0.14}. There is no signal loss, because the TRD particle identification was used whenever available and was not imposed as a strong requirement. \newline
\noindent These first studies already show that the inclusion of the TRD in the analysis largely improves the identification of $\rm J/\psi$. Further advancements in performance are ongoing. The Pb-Pb run in 2011 will increase the statistics especially because of the larger geometrical acceptance. In 2011 the TRD covers 10/18 of the azimuthal angle (7/18 in 2010). \newline
\noindent To significantly enrich the sample of quarkonia, the TRD detector will be used to select events with electron-positron pairs at the trigger level. For an overview of the TRD trigger see \cite{JKlein}.

\section{Conclusion and Outlook}\label{Section_ConclOutlook}
\noindent The ALICE Transition Radiation detector significantly extends the physics reach in ALICE enabling to study the conditions of the Quark-Gluon-Plasma with open heavy-flavour production, quarkonia, jets etc. The TRD provides good electron identification and allows to enhance rare probes by trigger contributions \cite{JKlein}. First physics results of electrons from heavy-flavour hadron decays were obtained using the one-dimensional likelihood on the total integrated charge measured in each chamber allowing to reduce the systematic uncertainty and to increase the measurable $p_{\rm t}$ range. More powerful particle identification methods with the TRD are under development. Including the TRD electron identification in the $\rm J/\psi$ studies in Pb-Pb collisions results in a reduced combinatorial background, better signal to background ratio as well as sig\-ni\-fic\-ance. The application of the TRD particle identification algorithm in further experimental data analyses is ongoing.





\bibliographystyle{model1-num-names}



\end{document}